# Real-projective-plane hybrid-order topological insulator realized in phononic crystals

Pengtao Lai[1†], Jien Wu[2†], Zhenhang Pu[3], Qiuyan Zhou[3], Jiuyang Lu[3], Hui Liu[1], Weiyin Deng[3*], Hua Cheng[1*], Shuqi Chen[1,4*], and Zhengyou Liu[3,5*]

[1]The Key Laboratory of Weak Light Nonlinear Photonics, Ministry of Education, Smart Sensing Interdisciplinary Science Center, School of Physics and TEDA Institute of Applied Physics, Nankai University, Tianjin 300071, China
[2]School of Physics and Optoelectronics, South China University of Technology, Guangzhou, Guangdong 510640, China
[3]Key Laboratory of Artificial Micro- and Nanostructures of Ministry of Education and School of Physics and Technology, Wuhan University, Wuhan 430072, China
[4]The Collaborative Innovation Center of Extreme Optics, Shanxi University, Taiyuan, Shanxi 030006, China
[5]Institute for Advanced Studies, Wuhan University, Wuhan 430072, China

†These authors contributed equally to this work.
*Corresponding author.
Emails: dengwy@whu.edu.cn; hcheng@nankai.edu.cn; schen@nankai.edu.cn; zyliu@whu.edu.cn

**The manifold of the fundamental domain of the Brillouin zone is always considered to be a torus. However, under the synthetic gauge field, the Brillouin manifold can be modified by the projective symmetries, resulting in unprecedented topological properties. Here, we realize a real-projective-plane hybrid-order topological insulator in a phononic crystal by introducing the $Z_2$ gauge field. Such insulator hosts two momentum-space non-symmorphic reflection symmetries, which change the Brillouin manifold from a torus to a real projective plane. These symmetries can simultaneously lead to Klein-bottle and quadrupole topologies in different bulk gaps. The non-symmorphic reflection symmetries on Brillouin real projective plane, edge states of Klein-bottle insulator, and corner states of quadrupole insulator are observed. These results evidence the hybrid-order topology on Brillouin manifold beyond the torus, and enrich the topological physics.**

Symmetry plays a key role in the classification of topological matter [1-3]. In the presence of gauge field, the algebraic structures of crystal symmetries can be projectively enriched and generate the so-called projective symmetries, opening up new avenues for topological physics [4,5]. A case in point is that the $\pi$ gauge flux can change the commute relation for reflection symmetries to the anti-commute one, forming the projective reflection symmetries and leading to the multipole insulator [6,7]. Among them, the quadrupole insulator is the two-dimensional second-order topological phase, and featured with the zero-dimensional corner states [8]. Recently, a unified discussion about the projective symmetry algebra based on $Z_2$ gauge field (0 and $\pi$ flux over the lattices) in the two dimensions has been constructed [9]. Under a $Z_2$ gauge field, the projective translation symmetry can give rise to the Möbius topological insulator [10,11], the projective mirror symmetry leads to mirror Chern insulator [12-14], and the projective PT symmetry can switch the spinless and spinful topological phases [15-18].

Interestingly, the $Z_2$ gauge field can modify the manifold of the fundamental domain of the Brillouin zone (BZ), giving rise to unprecedented topological phenomena. The Brillouin manifold is usually known as a torus. In the presence of $Z_2$ gauge field, the real-space reflection symmetry can be projectively represented, which generates the momentum-space non-symmorphic (MSNS) reflection symmetry that performs glide reflection in momentum space [19]. In two dimensions, a single MSNS refection symmetry can change the Brillouin manifold from a torus to a Klein bottle, and give rise to the Klein-bottle insulator (KBI), which possesses the first-order topology described by $Z_2$ invariant and hosts a pair of edge states with a nonlocal twist [20]. Abundant topologies on Brillouin Klein bottle are further discovered [21-25], including the first order and higher order. Very recently, it is proposed that a pair of MSNS reflection symmetries can reduce the Brillouin manifold from a torus to a real projective plane, and give rise to the second-order topological phase [26,27]. Experimental exploration of the intriguing topology on Brillouin real projective plane beyond the traditional paradigm is therefore necessary.

Here, we realize an acoustic real-projective-plane hybrid-order topological insulator (HOTI), which hosts a KBI phase with first-order topology and a quadrupole insulator (QI) phase with second-order topology in two bulk gaps. Owning to the macroscopic scale, phononic crystal (PC) for acoustic waves is a versatile platform to explore the abundant topological physics [28,29]. The synthetic $Z_2$ gauge field can be achieved by accurately designing the positive and negative couplings in the PC [30-33]. We first illustrate the MSNS reflection symmetries and the ensuing Brillouin real projective plane and topological phase diagram in the PC. We then observe the bulk properties on Brillouin real projective plane, KBI edge states, and QI corner states in a practical PC sample. The experimental results, in agreement with the theoretical ones, evidence the hybrid-order topology on Brillouin real projective plane.

Corresponding to the unit cell of lattice model shown in Fig. 1a, the unit cell of designed PC contains four same acoustic resonant cavities with the height $h = 30$ mm and width $r = 15$ mm, as illustrated in Fig. 1b. The lattice constant $a = 110.62$ mm. The intra-cell tubes with size $l = 3$ mm and inter-cell tubes with sizes $l_x = 4$ mm and $l_y = 5.5$ mm are constructed as the intra-cell and inter-cell couplings, which are at a distance of 10 mm to the top or bottom surfaces of cavities. The red and blue tubes act as the negative and positive couplings with $d_1 = 7$ mm and $d_2 = 8$ mm, as discussed in Supplementary S-I. With the cavities viewed as the lattice sites and the tubes acted as the couplings, the PC can be well mapped by the tight-binding model (Supplementary S-I). The Bloch Hamiltonian on the basis of the sublattices 1-4 can be written as

$$H(k_x, k_y) = \lambda_0(\Gamma_{01} + \Gamma_{13}) + \lambda_x(\cos k_x \Gamma_{31} + \sin k_x \Gamma_{32})$$
$$+\lambda_y(\cos k_y \Gamma_{10} + \sin k_y \Gamma_{20}), \quad (1)$$

where $\Gamma_{ij} = \tau_i \otimes \sigma_j$, $\tau_0$ ($\sigma_0$) is the $2 \times 2$ identity matrix, $\tau_i$ and $\sigma_i$ with $i = 1,2,3$ are Pauli matrices for the degrees of freedom within a unit cell. The lattice constant is set to unity, $\lambda_0$ is the intra-cell coupling, $\lambda_x$ and $\lambda_y$ are the inter-cell couplings along the $x$ and $y$ directions, respectively.

The PC possesses a pair of MSNS reflection symmetries $M_x^g = GM_x = \Gamma_{31}$ and $M_y^g = \Gamma_{10}$, where $M_x$ represents the normal reflection in the $x$ direction and $G = \Gamma_{30}$ is the gauge transformation. $M_x^g$ and $M_y^g$ satisfy the anti-commute relation $\{M_x^g, M_y^g\} = 0$. In momentum space, we have

$$M_y^g H(k_x, k_y)(M_y^g)^{-1} = H(k_x + \pi, -k_y), \tag{2}$$

$$M_x^g H(k_x, k_y)(M_x^g)^{-1} = H(-k_x, k_y + \pi). \tag{3}$$

To visualize these two symmetries, we illustrate the processes of the symmetric transformation. As shown in the left panel of Fig. 1c, for $M_y^g$ symmetry, the unit cell is reflected in the $y$ direction, then recovered by transforming $k_x$ into $k_x + \pi$. For $M_x^g$ symmetry, the unit cell is reflected in the $x$ direction, and needed to apply a gauge transformation $G$ by assigning a sign of $+1$ or $-1$ to the basis at each site, finally recovered by transforming $k_y$ to $k_y + \pi$, as illustrated in the right panel of Fig. 1c.

The MSNS reflection symmetries $M_x^g$ and $M_y^g$ reduce the BZ of PC to form a real projected plane manifold, which will act on the bulk dispersions shown in Figs. 1d and 1e. The first BZ is divided into 16 regions, and each region with the same color can be linked by the $M_x^g$, $M_y^g$ or $M_x^g M_y^g$ symmetry, as shown in Fig. 1e. The central region of the BZ ($[-\pi/2, \pi/2] \times [-\pi/2, \pi/2]$) consists of four regions marked by different colors, which contains the full information of the first BZ. Thus, we can define a reduced BZ of the central region as the fundamental domain of the first BZ. Interestingly, the boundaries of such reduced BZ are oppositely oriented via $M_x^g$ and $M_y^g$ symmetries, as shown by the red and blue arrows in Fig. 1e, and glued together to form a real projective plane manifold [34]. As a concrete example, we extract the iso-frequency contours of bulk dispersion at 5.5 kHz, as depicted by the black curves in Fig. 1e. All contours in the first BZ can be mapped from those in the real-projective-plane BZ. Under the $M_y^g$ ($M_x^g$) symmetry, A (B) point is transformed into D (C) point on the iso-frequency contours.

The MSNS reflection symmetries $M_x^g$ and $M_y^g$ can give rise to hybrid-order topology on Brillouin real projective plane. As shown in Fig. 1d, the four bulk bands can have three bulk gaps, in which the lower and upper gaps are the gaps of KBI and the middle gap is the gap of QI. The KBI possesses the first-order topology described by $Z_2$ invariants $w_x$ and $w_y$, while the QI is characterized by corner charge $Q_c$ that is defined by the quadrupole moment and edge polarizations, as discussed in detail in Supplementary S-II and S-III. For simplicity, we focus on the topological properties of the lower and middle gaps. Figure 1f shows the topological phase diagram determined by $(w_x, w_y; Q_c)$, in which $w_x$ and $w_y$ are calculated in the first band and $Q_c$ is for the lowest two bands. There exist four topological phases in the $l_x$-$l_y$ plane. Phases with $(1,0;0)$ and $(0,1;0)$ are the first-order topological phases with zero corner charge, while those with $(1,0;0.5)$ and $(0,1;0.5)$ are hybrid-order topological phases with nonzero first-order and second-order topological invariants simultaneously. The white line is the phase boundary of KBI with bulk gap closure, and the gray line is that of QI with edge gap closure, as shown in Supplementary S-IV.

We now experimentally reveal the MSNS reflection symmetries and Brillouin real projective plane in the PC. Figure 2a shows the PC sample fabricated by 3D printing technology with $(l_x, l_y) = (4 \text{ mm}, 5.5 \text{ mm})$, corresponding to the parameters labeled as the red star in Fig. 1f. The positive and negative hopping terms are realized by the coupling waveguides with different configurations. We calculate the Wannier bands $v_y(k_x)$ and $v_x(k_y)$ of the first bulk band by using its eigenfield, as shown in Fig. 2b. As $k_x$ ($k_y$) varies from $-\pi/a$ to 0, the Wannier band $v_y(k_x)$ ($v_x(k_y)$) passes through 0.5 with odd (even) times, which indicates the nontrivial (trivial) $Z_2$ invariant $w_x = 1$ ($w_y = 0$) in the $x$ ($y$) direction (Supplementary S-II). We also calculate the Wannier sector polarization $P_y^{v_x}$ and $P_x^{v_y}$ of the lowest two bands, as shown in Fig. 2c, and obtain nonzero quadrupole moment $q_{xy} = 2P_y^{v_x} P_x^{v_y} = 0.5$. The corner charge $Q_c$ can be calculated as $Q_c = p_x^e + p_y^e - q_{xy}$, where $p_x^e = 0.5$ and $p_y^e = 0.5$ are edge polarizations (Supplementary S-III), leading to $Q_c = 0.5$. So this PC is a HOTI with $(w_x, w_y; Q_c) = (1,0; 0.5)$. Figure 2d plots the calculated (white lines) and measured (color map) bulk dispersions along the Γ-M-X line. The calculated and measured iso-frequency contours at 5.5 kHz are displayed in Fig. 2e, denoting the existence of special MSNS reflection symmetry and ensuing real projective plane in momentum space. The measured and calculated iso-frequency contours at different frequencies are further shown in Supplementary S-V. The measured bulk dispersions are obtained by Fourier transforming the measured field, where the source is placed at the center of the PC sample. Therefore, the designed PC is a real-projected-plane HOTI.

According to the bulk-boundary correspondence, the real-projected-plane HOTI supports a pair of edge states with a nonlocal twist in the KBI gap. A PC ribbon is built to investigate the edge states along $x$ ($y$) direction, which has a periodic boundary condition in the $x$ ($y$) direction, and finite-size length in the $y$ ($x$) direction. Figures 3a and 3b show the calculated projected dispersions along $x$ and $y$ directions, respectively. It can be seen that the edge bands (colored lines) in the KBI gap only emerge in the $x$ direction, correspondence to the $Z_2$ invariants $(w_x, w_y) = (1,0)$. The color denotes the localization of edge state $\psi_e$ at the bottom (yellow) and top (blue) boundaries, which is defined as $d = \sum_{x \in b} |\psi_e(x)|^2 - \sum_{x \in t} |\psi_e(x)|^2$ and $b$ ($t$) represents the outermost two unit cells of the bottom (top) boundary. Due to the $M_y^g$ symmetry, the eigenfrequency at $k_x$ is the same as that at $k_x + \pi/a$, and their eigenfields are connected by the $y$ directional reflection symmetry, behaving as a nonlocal twist. For example, the frequency at $S_1$ ($k_x = -\pi/a$) is equal to that at $S_2$ ($k_x = 0$). The eigenfield distributions of edge modes at $S_1$ and $S_2$ are shown in Fig. 3c, where the color maps represent the acoustic pressures normalized by their maximum value. One can see that the fields of $S_1$ and $S_2$ are localized at top and bottom boundaries respectively, which satisfy $M_y^g \psi_{S_1} = \psi_{S_2}$ and verify the existence of MSNS reflection symmetry.

In experiments, we first place a source at the center of the top boundary and measure the response pressure of cavities in this boundary. After Fourier transforming the measured data, the projected dispersion is obtained and shown in the upper panel of Fig. 3d. Notably, only the edge modes near $k_x = \pm\pi/a$ are excited, revealing that the fields of these modes are mainly localized on the top boundary like $S_1$. The edge modes near $k_x = 0$ are excited by the source at the bottom boundary, revealing the localization of these edge modes on the bottom boundary like $S_2$, as shown in the lower panel of Fig. 3d. These results are consistent with the calculated ones in Fig. 3a, together revealing the edge states with a nonlocal twist.

Finally, we demonstrate the existence of corner states in the QI gap of the real-projected-plane HOTI, attributed to $Q_c = 0.5$. The square-shaped PC with $13 \times 13$ unit cells is depicted in Fig. 2a, and its schematic is displayed in Fig. 4a. Figure 4b shows the eigenfrequency spectrum, where the corner states (red dots) emerge in the QI gap. Inset displays four corner modes near the frequency 5.65 kHz. In the experiment, the source excites at each cavity and the response of acoustic pressure is measured at the same cavity. Figure 4c shows the measured pressure field distribution at frequency 5.65 kHz of corner states. One can see that the measured field is localized at four corners of the PC sample. We further measure the broadband response of acoustic pressures at corner ($C_1$), edge ($C_2$), and bulk ($C_3$), as shown in Fig. 4d, where the data are normalized by the maximum value. A huge resonant peak of pressure at corner is observed in the band gap, which is much larger than that at the edge and bulk, denoting the excellent ability of corner states in binding the acoustic pressure.

In conclusion, we have realized a real-projected-plane HOTI in a PC, where the BZ is reduced by a pair of MSNS reflection symmetries to form a real projective plane. The real-projected-plane HOTI simultaneously hosts that a KBI phase features a pair of edge states with a nonlocal twist in the lower gap and a QI phase characterizes nontrivial corner states in the middle gap. Our work represents an experimental extension of band topology on Brillouin manifold from a torus to a real projected plane. With the flexibility in achieving the synthetic gauge field, our system may serve as a basis for further exploration of nontrivial topological properties on three-dimensional Brillouin manifold beyond real projected plane.

**Methods**

**Details for simulation.** In this work, all the simulations are performed by the commercial COMSOL Multiphysics solver package, where the speed of sound is set as 341 m/s and the density of air is 1.3 kg/m$^3$. To calculate the Wannier bands in Fig. 2b, Wannier sector polarizations in Fig. 2c, and edge polarizations in Fig. S4, the eigenvectors are constructed by the normalized eigenfields which are extracted at the center of each cavity with the position of $h/4$.

**Details for experiment.** The experimental sample is fabricated by 3D printing technology with a resin thickness of 2 mm. Due to the large impedance mismatch with air, the resin boundaries can be regarded as the hard walls. The sample consists of $13 \times 13$ unit cells with sizes of around 1.4 m × 1.4 m. This big sample is divided into four parts, which are fabricated separately and assembled together, as shown in Fig. 2a. A hole with diameter 11.2 mm is perforated on the top side of cavities for excitation and detection. The holes are sealed with plugs when not in use. In experiments, the source with diameter 6.3 mm is placed at the top of cavities and the response signal is collected by the microphone, which is placed at the $h/4$ of cavities. Measured data in the frequency range 5 kHz - 6.5 kHz with frequency step 0.001 kHz is processed by the network analyzer (E5061B 5 Hz-500 MHz). The bulk band dispersions (Figs. 2d and 2e) and projected band dispersion (Fig. 3d) are obtained by Fourier transforming the corresponding measured fields, where a source is placed at the center of bulk and top/bottom boundary, respectively. For the measurement of corner states in Fig. 4c, the source and microphone are placed at the same cavity $i$, obtaining pressure response $P_i(f)$. To reduce the influence of varying excitation efficiencies in different cavities, response pressure in Fig. 4d is normalized by the sum of intensity over all frequencies $P_i(f)/\sum_f P_i(f)$.

## Acknowledgments

This work is supported by the National Key R&D Program of China (Grant Nos. 2022YFA1404501, 2022YFA1404900, and 2021YFA1400601), the National Natural Science Fund for Distinguished Young Scholars (Grant No. 11925403), the National Natural Science Foundation of China (Grant Nos. 12074128, 12122406, 12192253, and 12374409), and the Guangdong Basic and Applied Basic Research Foundation (Grant No. 2022B1515020102).

## Data availability

The data that support the plots within this paper and other findings of this study are available from the corresponding author upon reasonable request.

## Author contributions

All authors contributed extensively to the work presented in this paper.

**Competing financial interests**

The authors declare no competing financial interests.

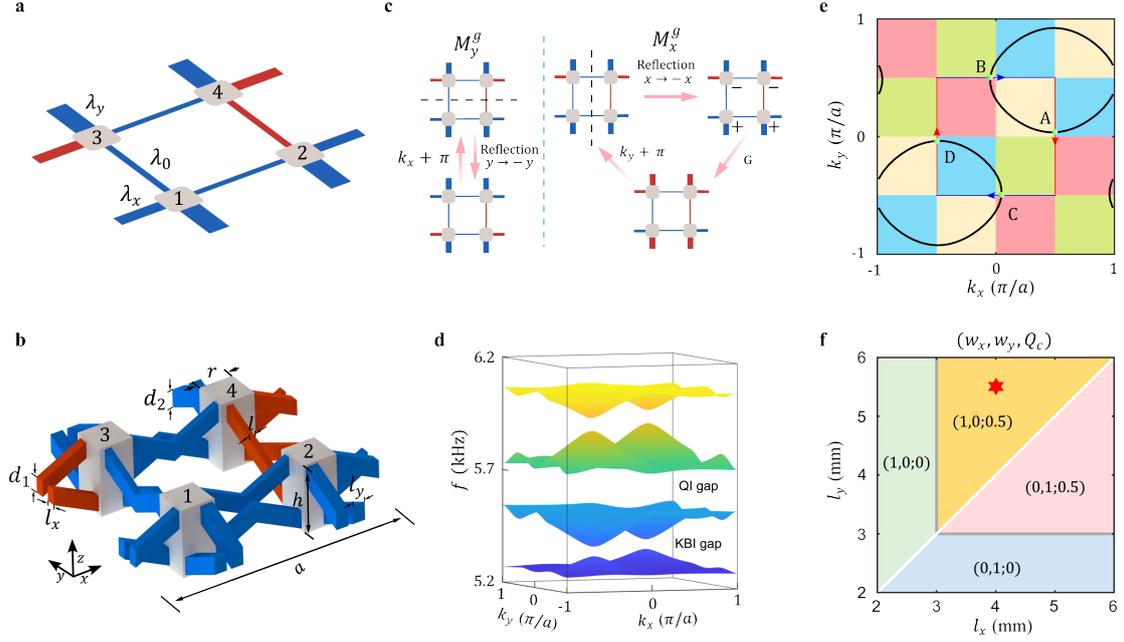

**Fig. 1 | Acoustic real-projective-plane HOTI protected by the MSNS reflection symmetries $M_x^g$ and $M_y^g$. a**, **b** Unit cells of tight-binding model and PC. Red (blue) tubes represent negative (positive) couplings. **c** Schematics of MSNS reflection symmetries. **d** Bulk dispersion of PC with KBI and QI gaps. **e** Iso-frequency contours (black curves) of PC at 5.5 kHz in the first BZ. The reduced BZ (enclosed by the red and blue arrows) can be regarded as a real projective plane. **f** Phase diagram determined by $Z_2$ invariants $(w_x, w_y)$ of KBI gap and corner charge $Q_c$ of QI gap in the $l_x$-$l_y$ plane. The white (gray) line denotes the phase boundary of KBI (QI). The red star represents the hybrid-order topological phase with specific parameters used in **b** and **d**.

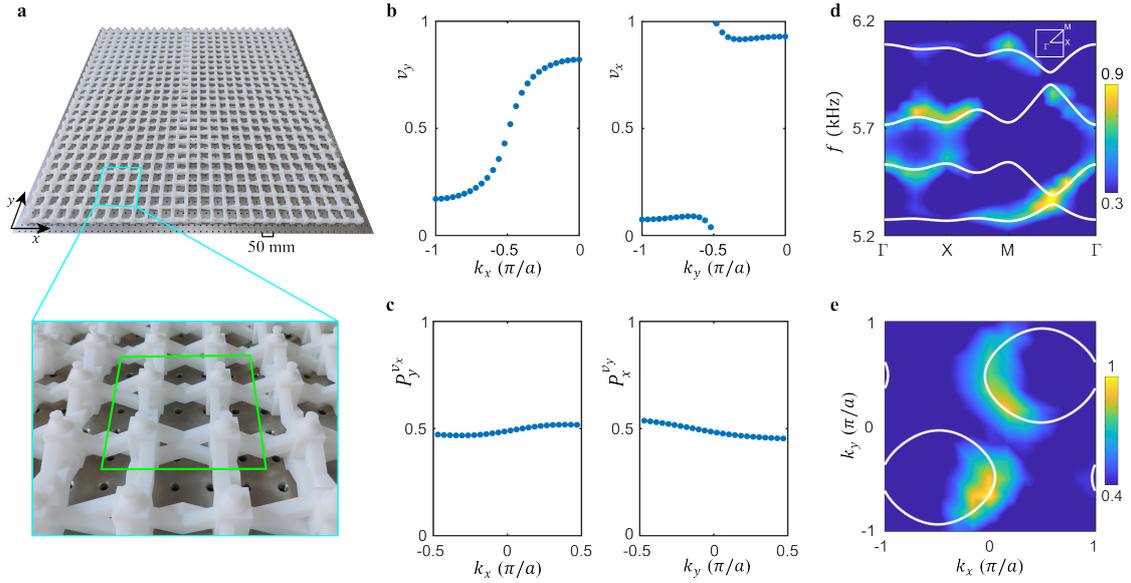

**Fig. 2 | Observation of bulk properties of acoustic real-projective-plane HOTI. a** Photograph of PC sample. The enlarged sample shows the detailed configuration of unit cell enclosed by green lines. **b** Calculated Wannier bands $v_y$ and $v_x$ of the first band of PC, which indicates the $Z_2$ invariants $(w_x, w_y) = (1, 0)$. **c** Calculated Wannier sector polarizations $P_y^{v_x}$ and $P_x^{v_y}$ of the lowest two bands of PC, revealing the quantized quadrupole moment. **d** Measured (color map) and calculated (white lines) dispersions along the Γ-X-M-Γ lines. **e** Measured and calculated iso-frequency contours at 5.5 kHz.

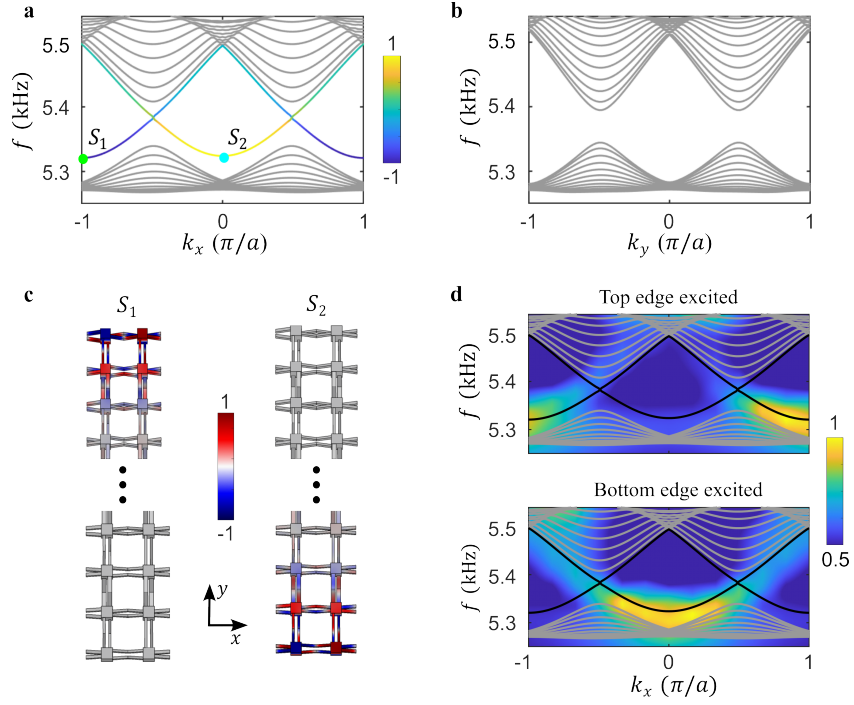

**Fig. 3 | Observation of edge states in the KBI gap. a** Calculated projected dispersions along the $x$ direction. The colored lines represent a pair of edge states with a nonlocal twist in the KBI gap, and the color denotes the degree of localization at the top (blue) and bottom (yellow) boundaries. **b** Calculated projected dispersion along the $y$ direction. There is no edge state in the KBI gap. **c** Eigenfield distributions of edge modes at $S_1$ ($k_x = -\pi/a$) and $S_2$ ($k_x = 0$) points marked in **a**, which are connected by $M_y^g$ symmetry. **d** Measured edge dispersions (color maps) excited at the centers of the top and bottom boundaries, respectively. The black lines are the calculated edge dispersions in **a**.

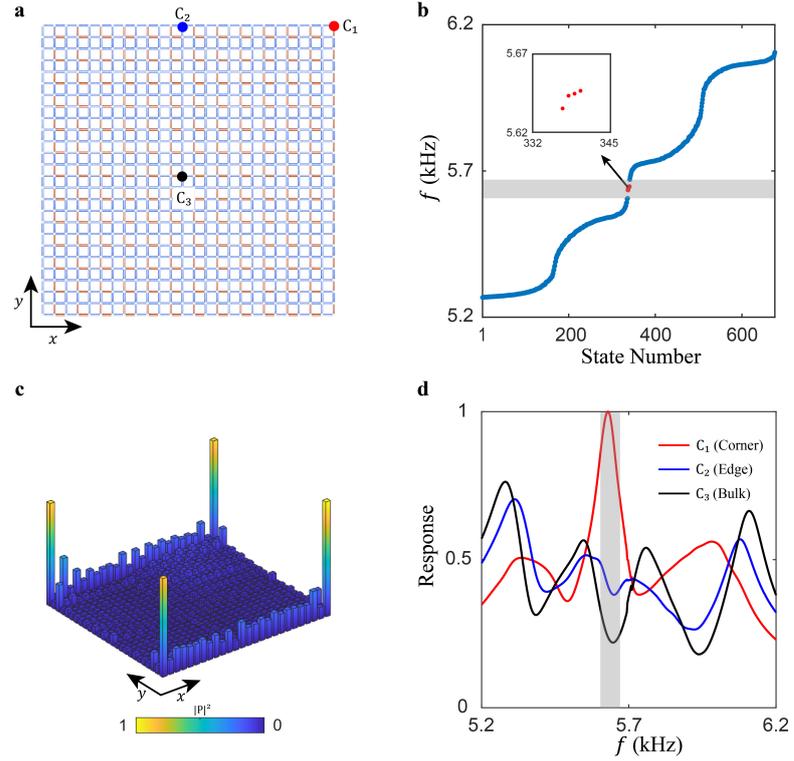

**Fig. 4 | Observation of corner states in the QI gap. a** Schematic of square-shaped PC sample. **b** Eigenfrequency spectrum. The four red dots denote the corner modes in the QI gap. Inset: enlarged region for corner states. **c** Measured pressure field distribution at 5.65 kHz, demonstrating the existence of corner states. **d** Measured response spectra at the positions $C_1$, $C_2$, and $C_3$, which are normalized by the maximum value.